\documentclass[12pt]{article}
\usepackage {amsmath} 
\usepackage{graphicx}
\usepackage{url}

\newcommand {\eqrefa}[1]{(\ref {#1})}

\begin{document}

\begin{titlepage}

\vspace*{20mm}
\begin{center}
{\Large {\bf The electrodynamic 2-body problem \\
and the origin of quantum mechanics\footnote{This is a pre-press version. The original publication is available at doi:\\
10.1023/B:FOOP.0000034223.58332.d4
}} \\}

\vspace*{15mm}
\vspace*{1mm}
{C.K. Raju \footnote{c\_k\_raju@vsnl.net} }

\vspace*{1cm}

{\it Centre for Computer Science\\
MCRP University, Gyantantra Parisar\\
222, M.P. Nagar Zone I, Bhopal 462 011, India}

\vspace*{0.4cm}

\vspace*{1cm}
\end{center}
\end{titlepage}

\begin {abstract}

We numerically solve the functional differential equations (FDE's) 
of 2-particle electrodynamics, using the full electrodynamic force 
obtained from the retarded Lienard-Wiechert potentials and the 
Lorentz force law.  In contrast, the usual formulation uses only 
the Coulomb force (scalar potential), reducing the electrodynamic 
2-body problem to a system of ordinary differential equations (ODE's). 
The ODE formulation is mathematically suspect since FDE's and ODE's 
are known to be incompatible; however, the Coulomb approximation to 
the full electrodynamic force has been believed to be adequate for 
physics. We can now test this long-standing belief by comparing the 
FDE solution with the ODE solution, in the historically interesting 
case of the classical hydrogen atom. The solutions differ.

A key qualitative difference is that the full force involves 
a `delay' torque.  Our existing code is inadequate to calculate 
the detailed interaction of the delay torque with radiative damping. 
However, a symbolic calculation provides conditions under which the 
delay torque approximately balances (3rd order) radiative damping. 
Thus, further investigations are required, and it was prematurely 
concluded that radiative damping makes the classical hydrogen atom 
unstable. Solutions of FDE's naturally exhibit an \textit{infinite} 
spectrum of \textit{discrete} frequencies.  The conclusion is that 
(a)~the Coulomb force is \textit{not} a valid approximation to the 
full electrodynamic force, so that (b)~the \textit{n}-body 
interaction needs to be reformulated in various current contexts 
such as molecular dynamics.

\end{abstract}

\textbf{pacs: } {03.50.De, 03.30.+p, 02.30.Ks, 02.90.+p, 
87.15.Aa, 31.15.-p, 03.65.Sq, 03.65.Ta, 95.35.+d}

\textbf{keywords: } {many-body problem, protein dynamics, functional 
differential equations, relativistic many-body problem,  
interpretation of quantum mechanics.} 


\pagebreak

\section {INTRODUCTION}

\subsection {Aim}

\noindent This author had earlier$^{(1)}$ proposed a new model 
of time evolution in physics using mixed-type functional differential 
equations (FDE's), with a tilt in the arrow of time. This paper sets 
aside the notion of a `tilt', and takes up only the FDE's of retarded 
electrodynamics. The retarded case already explicitly incorporates 
certain subtle mathematical features of electrodynamics and relativity 
noticed by Poincar\'{e}, but overlooked by Einstein and subsequent  
researchers. To bring out these subtleties, this paper reports on 
a numerically computed solution of FDE's of the 2-body problem of 
classical retarded electrodynamics.$^{(2)}$ 

The use of the full (retarded) electrodynamic 2-particle force leads 
to the formulation of the electrodynamic 2-body problem as a system 
of FDE's that have \textit{not} actually been solved earlier, 
numerically or otherwise, despite some sporadic attempts in 
simplified situations.$^{(3)}$ In the absence of a systematic 
way to solve these FDE's, a widely used alternative  has been to 
approximate the full electrodynamic 2-particle force by the Coulomb 
force. This reformulates the electrodynamic 2-body problem as an 
easier system of ODE's, which can be numerically solved with 
exactly the same numerical techniques that are used for the 
ODE's of the classical 2-body problem of Newtonian gravitation. 
This alternative ODE formulation of the 2-particle electrodynamic 
interaction is incorporated, for example, in models of protein 
dynamics$^{(4)}$ underlying current software such as {\sc charmm}, 
{\sc wasser}, {\sc amber} etc.

This alternative ODE formulation is, however, mathematically suspect, 
for it is known that solutions of FDE's may exhibit qualitative 
features impossible for solutions of ODE's. On the other hand, 
it is believed that, from the viewpoint of physics, the Coulomb 
force is an adequate approximation to the full electrodynamic force.  

We can now put this long-standing belief to test: our numerical 
solution of the full-force FDE's enables us compare the two 
solutions in the historically interesting context of the 
classical hydrogen atom.  

\subsection {The full electrodynamic force}

\noindent In classical electrodynamics, the force between moving 
charges is given by the Lienard-Wiechert potentials combined with 
the Heaviside-Lorentz force law. The scalar and vector (retarded) 
Lienard-Wiechert potentials, are given by the expressions$^{(5)}$ 

\begin{equation} 
V ( {\bf r} ,  \:  t )  \: = \: {1 \over 4 \pi \epsilon_0 }  \:  
{ qc \over ( R c -  \:  {\bf R \cdot v} ) }  \: ,  
 \:  \:  \:  \:  
{\bf A} ( {\bf r} ,  \: t )  \: = \:  { {\bf v} \over  c^2}  
V ( {\bf r} ,  \:  t).
\label{L-W}
\end{equation}

\noindent One now computes the fields $\bf E$, and $\bf B$ 
by computing

\begin{equation} 
{\bf E}  \: = \:  - \nabla V \:  - \:   { \partial {\bf A} \over 
\partial t } \: , 
\quad
{\bf B}  \: = \:  \nabla  \:  \times  \:  {\bf A} .
\label{L-B}
\end{equation}

\noindent The expressions for these fields are:$^{(6)}$ 

\begin{equation} 
{\bf E}  \: = \:  { q \over 4 \pi \epsilon_0 }  \cdot   
{R  \over ( {\bf R \cdot u} )^3 }  
 \:  \: [ { \bf u} \:  (c^2 - v^2 )   
 \:  +  \:  { \bf R}  \: \times  \:  ( { \bf u} \:  \times  
 \:  {\bf a} ) ] ,
\label{em field}
\end{equation}

\begin {equation}
{\bf B}  \: = \:  {1 \over c}  {\bf \hat R} \times {\bf E} .
\end{equation}

When these expressions for the fields $\bf E$, and $\bf B$ 
are substituted into the Heaviside-Lorentz force law,$^{(7)}$ 

\begin{equation} 
{\bf F} \: = \:  q_1 \, \left ( {\bf E}  \: +  \:  {\bf v}_1    
\times {\bf B} \, \right ) ,
\label{H-L}
\end{equation}

\noindent the force on a charge $q_1$ moving with velocity 
$v_1$ is given by the expression:$^{(8)}$ 

\begin{eqnarray}
{\bf F}  \: & = & \:  { q q_1   \over 4 \pi \epsilon_0}
 {R \over ( \bf R \cdot u )^3 }  
 \left \{ \left [ ( c^2 - v^2 )  {\bf u} +   
{\bf R}  \times  ( u \times a ) \right ] \nonumber  \right. \\
 & & {} +  {{\bf v}_1 \over c} \times  \left. \left [ {\bf \hat R}   
\times [( c^2 - v^2 )  {\bf u}   +   
 {\bf R } \times ( {\bf u} \times {\bf a} ) ]   
 \right ]   
\right \} .
 \label {em-force}
\end{eqnarray}

Here, charge $q_1$ is located at ${\bf r} (t)$ at time $t$, 
while the position of the other charge $q$ at time $t$ is given 
by ${\bf w} (t)$, and

\begin{equation} 
{\bf R}  \: = \:   {\bf r} ( t)  \: - \: {\bf w} ( t_r ) .
\end{equation}

\noindent In the above expression, $t_r$ is the retarded 
time (the time at which the backward null cone with vertex at 
${\bf r} (t)$ meets the world line ${\bf w} (t)$ of the other charge), 
and is given implicitly by the equation

\begin{equation} 
||  {\bf r} (t) - {\bf w} (t_r) ||  \: = \:  c ( t - t_r ) .
\end{equation}

\noindent Further, 

\begin{equation} 
{\bf u} \: = \:  c { {\bf R}  \over  R}   \: - \: 
{\bf v} \: = \:  c {\bf \hat R} 
 - {\bf v} ,
\end{equation}

\noindent and it is understood that 
${\bf v}  \: = \:  {\bf \dot w} ( t_r )$ and 
${\bf a} \: = \:  {\bf \ddot w} ( t_r )$ are the velocity 
and acceleration of the charge $q$ at the \textit{retarded time} 
$t_r$.  A similar expression gives the force exerted by the charge 
$q_1$ on the charge $q$. This leads to the formulation of 
the 2-body problem as a system of FDE's. 

\subsection {The Coulomb approximation}

\noindent An alternative formulation approximates the full force 
by the Coulomb force. This approximation has been justified as 
follows. In the Coulomb gauge, $\nabla \cdot {\bf A}  \: = \:  0$, 
the scalar potential $V$ satisfies the Poisson equation. 
\textit{Neglecting the vector potential}, the force between the 
two particles (i.e., the force due to this scalar potential) is 
just the Coulomb force of electrostatics, 

\begin{equation} 
{\bf F}  \: = \:  { q q_1 \over r^3} {\bf r} ,
\end{equation}

\noindent $\bf r$ being the instantaneous separation of the 
two particles. This leads to the formulation of the electrodynamic 
2-body problem as a system of ODE's, similar to the 2-particle ODE's 
of Newtonian gravitation. 

\subsection {The difficulty: FDEs vs ODEs}

\noindent However, the vector potential cannot be so easily 
neglected---the possible justification suggested above involves 
an all too facile analogy between the scalar Lienard-Wiechert 
potential and electrostatics. In the language of a text, 

\begin{quote}
Don't be fooled, though---unlike electrostatics, $V$  by 
itself doesn't tell you $\bf E$, you need to know $\bf A$ as 
well.\ldots $V$ by itself is not a physically measurable 
quantity---all the man in the moon can measure is $\bf E$, and 
that involves $\bf A$ as well. Somehow it is built into the 
vector potential, in the Coulomb gauge, that whereas $V$ 
instantaneously reflects all changes\ldots $\bf E$ will change 
only after sufficient time has elapsed for the `news' 
to arrive.$^{(9)}$ 
\end{quote}

\noindent Briefly, the Coulomb force (based on $V$) acts 
instantaneously, while the full force (based on $\bf E $) involves a 
delay (assuming retarded potentials).

Neglecting the vector potential, hence the delay, corresponds 
mathematically to approximating FDE's by ODE's. This is suspect 
since it is known that FDE's are fundamentally different from 
ODE's, and that solutions of FDE's can have qualitative features 
impossible for solutions of ODE's. The physical consequences of 
these differences are explained at length in the author's previous 
book.$^{(10)}$ Two differences of immediate concern are the 
following. 

(a) Given a system of ODE's of the form

\begin{equation} 
{ \bf \dot x} ( t)  \: = \:  g ( t,  \:  {\bf x} ( t ) ) 
 \label {ODE}
\end{equation}

\noindent a unique solution can be obtained by prescribing the 
initial values of $\bf{x}$ at a single point of time, say 
${\bf x} ( 0 ) $. However, given a system of FDE 's of the form

\begin{equation} 
{\bf \dot x} ( t )  \: = \:  f \left ( t,  \: { \bf x} ( t ) ,  \: 
{\bf x} ( t - t_r )  \right )  
 \label {FDE}
\end{equation}

\noindent and assuming these FDE's to be retarded (i.e., 
$t_r > 0$), this is no longer true. For example, consider the FDE

\begin{equation} 
\dot x ( t)  \: =  \:  x ( t - {\pi \over 2} ) . 
\label{cos}
\end{equation}

\noindent Clearly, $x ( t)  \: = \:  \sin ( t )$ and 
$x ( t )  \: = \: \cos ( t )$ are solutions, and since the 
equation is linear, any linear combination 
$x ( t )  \: = \:  a \sin ( t )  \: + \: b \cos ( t) $ is 
also a solution of \eqrefa {cos}, for arbitrary constants $a$ and $b$, 
and the values of both $a$ and $b$  obviously cannot be fixed 
from a knowledge of a single quantity $x ( 0 ) $.  

In fact, from the mathematical theory of FDE's,$^{(11)}$ it is known 
that prescribing the initial values of even an infinite number of 
derivatives, $x (0),~\dot x(0),~\ddot x (0)~ \ldots$ 
is not adequate. To obtain a unique solution of the retarded FDE 
\eqrefa {FDE} we are required to prescribe the values of $x(t)$ over an 
entire interval  $\left [ 0,  \:  - t_r \right ] $, in the form of an 
\textit{initial function} $\phi$. That is, to solve retarded FDE's 
one needs to prescribe \textit{past} data  rather than instantaneous 
or `initial'(or final) data. 

(b)~Since the speed of light is so large, in many actual contexts, 
the delay $t_r$ is likely to be extremely small 
($t_r  <  10^{-18}$s, for the classical hydrogen atom). Under these 
circumstances, it is tempting to approximate a FDE  of the form 
\eqrefa {FDE} by an ODE of the form \eqrefa {ODE} by arguing that if the 
delay $t_r$ is small, then the values of $x ( t - t_r ) $ can 
be approximately obtained from the values of $x ( 0 ) $, 
$\dot x (0)$, $\ddot x ( 0 ) , \dddot x (0), \ldots$, by means 
of a `Taylor' expansion. Such a procedure, however, is known 
to be erroneous because of fundamental \textit{qualitative} 
differences between solutions of FDE's and ODE's. For example, 
unlike an ODE, which can be solved either forward or backward in 
time, a retarded FDE cannot generally be solved backward in time. 
Fig. 1 reproduced from Ref. 1 depicts `phase collapse'---three 
forward solutions of an FDE merge into one solution, so that a 
unique backward solution is impossible from data prescribed 
on the interval [1, 2]. Intersecting trajectories in phase space 
is a phenomenon impossible with ODE's, so that the basic classical 
mechanics requirement of a phase flow breaks down.$^{(12)}$ 
[The origin of this time-asymmetry is  retrospectively obvious on 
physical grounds, since the force \eqrefa  {em-force} has destroyed the 
underlying time symmetry of electrodynamics, by 
assuming retardation.] 

\begin{figure}[t]
\centerline {\scalebox{1.2}[1.2]{\includegraphics 
		[width = 18pc] {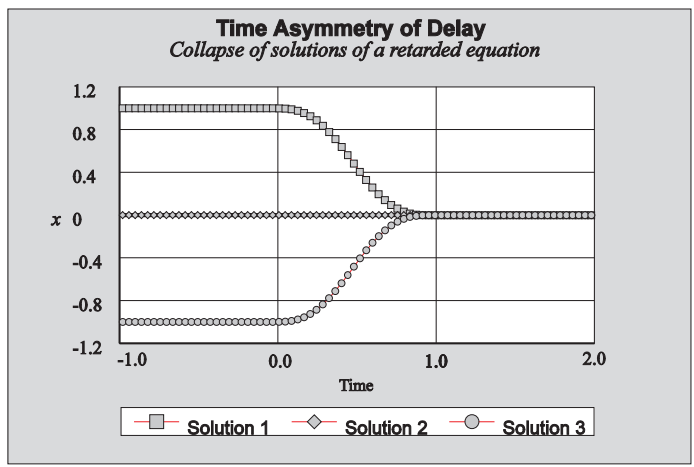}}}
\caption {\small{\it{The figure shows three different solutions of a retarded 
FDE $\dot x (t) \: = \: b(t) x (t-1)$, for a suitable choice of the 
function $b(t)$. Three different past histories prescribed for 
$t \leq 0$ lead to three different solutions all of which coincide 
for $t \geq 1$. Such a phase collapse is impossible with ODE where 
trajectories in phase space can never intersect. Because of this 
phase collapse, FDE, unlike ODE,  cannot be solved backward, from 
prescribed future data.}}}

\label {collapse}
\end{figure}

Thus, on grounds of the known mathematical differences between 
FDE's and ODE's, it is reasonable to doubt, a priori, that the 
full electrodynamic force \eqrefa  {em-force} can be validly 
approximated by the Coulomb force, by neglecting the vector 
potential. 

A good way to settle this doubt is to put the matter to test by 
comparing the Coulomb-force solution with the solutions obtained 
using the full electrodynamic force. This requires a solution of 
the 2-body problem with the full electrodynamic force, and we 
accordingly proceed to calculate such a solution.  

\section {SOLVING FDE's}

\subsection {Method 1}

\noindent The mathematical features of FDE's briefly recapitulated 
above suggest that retarded electrodynamics involves a `paradigm 
shift' from `classical mechanics', for we now need past 
data rather than instantaneous data.  In a debate on this question 
at Groningen, H. D. Zeh argued against any such paradigm shift. Zeh 
maintained that there was no need for past data, since data prescribed 
on a spacelike hypersurface (corresponding to an instant in spacetime) 
was adequate to solve the system of Maxwell's partial differential 
equations (PDE's) together with the Heaviside-Lorentz equations of 
motion for each particle. 

Given all fields on a spacelike hypersurface, one can evolve them 
forward in time for a small region. \textit{Given all fields}, in the 
vicinity of a single particle, its equations of motion reduce to a 
system of ODE's which can be solved in the usual way. We will call 
this method~1. We note that it implicitly involves the simultaneous 
solution of a \textit{coupled} system of PDE's and ODE's.

\subsection {Method 2}

\begin{figure}[t]
\centerline {\includegraphics 
		[width = 12pc] {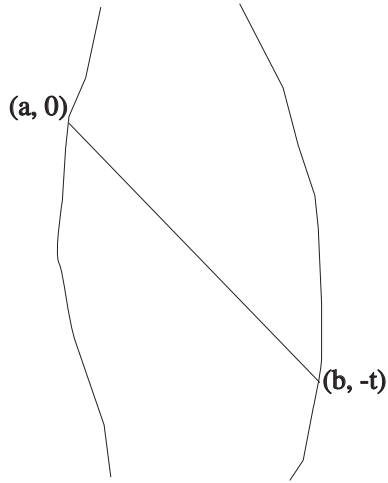}}
\caption{\small{\it{To calculate the force acting on particle 1 at $(a, 0)$, one draws the 
backward null cone with vertex at $(a, 0)$. If this intersects the world 
line of the second particle at $(b, -t)$, one calculates the 
Lienard-Wiechert potential at $(a, 0)$, due to the motion of 
particle 2 at $(b, -t)$.  Thus, determination of the present force on 
particle 1 requires a knowledge of the past motion of particle 2. }}}
\label {twobodyforce}
\end{figure}

\noindent In contrast to method~1, the FDE formulation of the 
electrodynamic 2-body problem may be geometrically visualised 
as follows.$^{(13)}$ Assuming retarded potentials, the force 
on particle~1 at a given spacetime point $(a,  \: 0)$ is 
evaluated as follows. One constructs the backward null cone 
with vertex at $(a,  \: 0)$, and  determines where it intersects 
the world-line of particle~2, at a point $(b,  \:  t) $, say 
(see Fig.~\ref{twobodyforce}).  Given the world-line of the particle 2 in a neighborhood of $(b,  \:  t)$ one evaluates 
the resulting Lienard-Wiechert potential at $(a,  \: 0)$, 
and uses this retarded potential to calculate the force on 
particle~1 at $(a,  \: 0)$. This is the force given by 
\eqrefa  {em-force}. The force on particle~2 at any point is 
calculated similarly. Clearly, we can solve for the motion of 
either particle, only if we are given the appropriate portions 
of the \textit{past} world line of the other particle. We will 
call this method~2.

\subsection {Relating the two methods}

\noindent The Groningen debate brought out the following difficulty. 
\textit{Both} the above methods seem to have the \textit{same} 
underlying physical principles (Newton's laws of motion, possibly 
in a generalised form suitable for special relativity + Maxwell's 
equations).  How, then, can these principles admit fundamentally 
incompatible interpretations of instantaneity and history-dependence? 
\subsection {The need for past data}

\begin{figure}[t]
\centerline {\includegraphics 
		[width = 14pc]{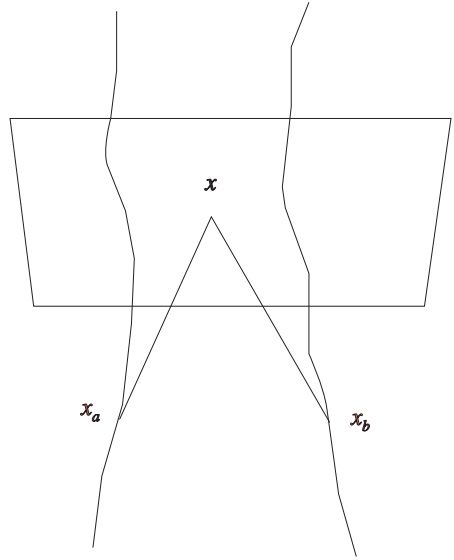} }
\caption {\small{\it{Method 1 seems to requires only instantaneous data corresponding to 
fields on a Cauchy hypersurface. Assuming only two particles and 
retarded propagators, the fields at any point $x$ on the hypersurface 
relate to past particle motions at $x_a$ and $x_b$. As 
$x$ runs over the hypersurface, the points $x_a$ and $x_b$ 
will, in general, cover the entire past. Thus, in the retarded 
case, initial data for the fields is the same as past data for the 
world lines of the particles.  }}}
\label {relating1and2}
\end{figure}

\noindent This issue can be readily resolved as follows. If retarded 
propagators (i.e., Green functions, or fundamental solutions of the 
wave equation, or retarded Lienard-Wiechert potentials) are assumed, 
then the field at any point $x$  relates to particle movements in 
the past at the points $x_a$ and $x_b$ on the respective 
world lines of  particles $a$ and $b$, where the backward null 
cone from $x$ respectively meets the world lines of the two particles(Fig.~\ref{relating1and2}). 
Thus, to know the field at a 
point \textit{x}  we should know the particle world lines 
at and around the \textit{past }points $x_a$, and $x_b$. 
If we do this for every point $x$ on a spacelike hypersurface, 
the points $x_a$, and $x_b$ will, in general, cover the 
entire past world lines of the two particles.  Thus, in the 2-body 
context, given the assumption of retardation, prescribing the fields 
on a spacelike hypersurface is really equivalent to prescribing the 
\textit{entire} past histories of the two particles. That is,

\begin{displaymath} 
\textit {instantaneous data for e.m. fields = past data for world 
lines of particles}
\end{displaymath}
That is, the PDE+ODE method 1 only \textit{hides} the underlying 
history-dependence of electrodynamics. (The above remarks need to 
be appropriately modified if, instead of assuming retardation, 
one assumes advanced or mixed-type propagators. For example, if 
we use advanced propagators, then `anticipation' should be used 
in place of `history-dependence', etc.) 

\subsection {FDE vs ODE+PDE}
\noindent While \textit{both} methods require past data on the motion 
of the two particles, the intuitive schema underlying method~1 is 
currently inconvenient for the actual process of obtaining a solution. 
Formally speaking, there is no well-known existence and uniqueness 
theorem for a \textit{coupled} system of PDE's+ODE's. [Separate 
existence theorems are, of course,  known for PDE's (Maxwell's 
equations, in this case), and for ODE's (to which the particle 
equations of motion reduce, \textit{if} all fields are given).]  
Neither is there any well known numerical algorithm which converts 
the intuitive method of iteratively solving coupled PDE's + ODE's 
into an actual process of calculation.$^{(14)}$ Both formal proof 
and a numerical scheme can very likely be developed without much 
difficulty. However, neither is available as of now, so method~1 
cannot, as of now, be used to obtain a solution of the full-force 
electrodynamic 2-body problem. 

In contrast, for FDE's there already is a formal existence and 
uniqueness theorem from past data.$^{(15)}$ Further, there are 
well known numerical algorithms and tested computer programs 
available$^{(16)}$ (though they do not~have all of the most 
desirable numerical characteristics, and the current code has 
not been formally proved to be error free). Accordingly, method~2 
is currently the method of choice. 

Incidentally, the PDE+ODE method~1 seems to need data on the 
\textit{entire} past trajectories of the two particles. This is 
\textit{more} information than is usually required for the FDE 
method~2.  [Because of the numerical stiffness of the underlying 
equations, in practice, one is able to numerically solve the 
classical hydrogen atom with method~2 for only short time periods, 
of the order of a femto second ($10^{-15}$~s), for which only a 
very short portion of the past history is needed.] That is, 
method~1, despite its appearance of preserving instantaneity and 
not needing any information from the past,  actually seems to 
need \textit{more} information about the 
past than method~2! A very careful analysis of the method would 
presumably show that the solution by method~1 actually uses 
information from only that part of the Cauchy 
hypersurface within the Cauchy horizon, so that information across 
the entire hypersurface is not needed, and the two methods are 
really equivalent. At present, however, that is still a conjecture.

\section {SOLUTION OF THE ELECTRODYNAMIC TWO-BODY PROBLEM}

\subsection {Prescription of past history and discontinuities}

\noindent \textit{How} should the past history of the two particles 
be prescribed? Existing physics provides no guidelines to help answer 
this question: exactly like fields on a spacelike hypersurface in 
method~1, it permits us to prescribe the past particle motions more 
or less arbitrarily.

This has two consequences worth noting. Thus, (1)~mathematically, 
the past data for a FDE is \textit{not} required to be a solution of 
the FDE, and (2)~the mathematical theory of FDE's tells us that a 
discontinuity may well develop at the initial point where the 
prescribed past data joins with the solution of the FDE. 

From a physical point of view, this past motion may be regarded as 
a bound or constrained motion---constrained, perhaps, by additional 
mechanical forces---and the discontinuity at the initial point may 
be attributed to the sudden removal of the additional constraint at 
$t=0$. These discontinuities propagate downstream. While it is known 
from the general theory$^{(17)}$ that the discontinuities of a 
retarded FDE are typically smoothened out over time, i.e., that 
they move over to successively higher derivatives, we must 
recognize that the problem we are solving is (except in 1~dimension) 
technically a neutral$^{(18)}$ FDE (even though only retarded 
Lienard-Wiechert potentials are being used), so that no such 
smoothing need take place, and the discontinuities may persist. 

The discontinuity creates a mathematical difficulty: for, in 
classical electrodynamics, the equations of motion are formulated 
assuming that the particle trajectories are at least thrice 
continuously differentiable ($C^3$). The numerical solution of 
the FDE by higher-order Runge-Kutta methods also assumes a 
similarly high level of smoothness, which assumption is not 
justified if the past history is prescribed arbitrarily. 

This problem of discontinuities obviously is a suitable topic for 
further research, and in the sequel we shall set it aside and rely 
on the methods for handling discontinuities that are already 
incorporated into existing computer codes like \textsc{retard} and 
\textsc{archi}$^{(19)}$, which handle discontinuities for example 
by switching to low-order polynomial extrapolation or 
predictor-corrector methods near a discontinuity. Both these 
programs are so written that they permit a discontinuity even in 
the function (positions, velocities) at the initial point, though 
we can always arrange initial conditions so that there is a 
discontinuity at most in the derivative (acceleration) at the 
initial point.  We will prescribe past and initial data in such 
a way that discontinuities are confined to the acceleration and 
higher derivatives---though, like `quantum jumps', it is not at 
present clear whether these are the only sorts of discontinuities 
that are `physically acceptable'.  The \textsc{archi} program 
enables the tracking of such discontinuities, classified as `hard' 
or `soft' according to the theory of Will\'{e} and 
Baker.$^{(20)}$ Hard discontinuities are those that propagate 
instantaneously between components, while soft discontinuities 
are those that propagate over time. In the present context, since 
there are no advanced interactions, all discontinuities are soft, 
except for discontinuities between components (velocity, 
acceleration) representing derivatives. (Further, the 
\textsc{archi} program has certain refinements---such as the use of 
a cyclic queue to store past history---that are not immediately 
relevant.)

\subsection {The classical hydrogen atom}

\noindent To return to the physics, since our aim is to test how 
well the Coulomb force approximates the full electrodynamic force, 
let us consider the case of the classical hydrogen atom, without 
radiative damping. Suppose an electron and proton are for 
$t \leq 0$ constrained to rotate rigidly in a classical circular 
2-body orbit, calculated using the Coulomb force. What will happen 
when this constraint is removed  at $t=0$? With the past history 
prescribed in this way, we now have a complete problem that can 
be solved by method~2.

\subsection {Equations of motion}

\noindent That is, we take the non-relativistic equations of motion 
to be given by the Heaviside-Lorentz force law for each 
particle:$^{(21)}$

\begin{equation}
m {d {\bf v}  \over  dt}  \: = \:  q \, \left ( { \bf E}  \: 
+  \:   { \bf v}  \times {\bf B}  \,  \right )  \:  
\left [ + {2 \over 3}  \, 
{q^2 \over 4 \pi \epsilon_0 c^3} {d^2 { \bf v}  \over  d t^2 }   
\right ]  .
\label {eom}
\end{equation}

\noindent The radiative damping term given by square brackets in 
\eqrefa  {eom} has been dropped, since it is not immediately relevant 
to our purpose of deciding whether the electrodynamic force 
\eqrefa  {em-force} can be approximated by the Coulomb force. (We know 
that, in the absence of radiation damping, with the Coulomb force, 
the initial `Keplerian' orbit should remain stable.) We also take 
the mass to be constant though, relativistically, only the proper 
mass is constant. This assumption, again, is appropriate to the 
comparison we wish to make between the Coulomb force and the full 
electrodynamic force. 

\subsection {Notation for the explicit system of equations}

\noindent A slight change of notation is helpful for the explicit 
calculation. We let

\begin{equation} 
\kappa  \: = \:  {q_1 \,  q_2  \over 4 \pi \epsilon_0  m_1 }, 
 \quad
\mu  \: = \:  {m_1 \over m_2}  .
\end{equation}

\noindent The mass ratio $\mu$ is dimensionless, while $\kappa$ 
has dimensions of $L^3 T^{-2}$ (=length in units in which 
$c=1$). Denote the 3-dimensional trajectories of the two particles 
by ${\bf r}_1 (t) , \:  {\bf r}_2 ( t )$. The delays 
$\tau ,  \:  \bar \tau$ , and the vectors ${\bf R}_1 ,  \: 
{\bf R}_2$ are defined by the simultaneous equations

\begin{equation} 
c^2  \:  \: ~\tau ^2  \: = \:  R_1^2 , 
\quad
{\bf R}_1  \: = \:  {\bf r}_2 ( t )  
- {\bf r}_1 ( t - \tau ), 
\quad 
R_1  \: = \:  || {\bf R}_1  ||  , 
\end{equation}

\begin{equation} 
c^2   \:  {\bar \tau }^2  \: = \:  R_2^2 , 
\quad
{\bf R}_2  \: = \:  {\bf r }_1 ( t )  
- {\bf r}_2 ( t - \bar \tau ),  
\quad  
R_2  \: = \:   || {\bf R}_2  || .
\end{equation}

\noindent Since here we consider only retarded solutions, for which 
$\tau , \bar \tau > 0 $ we need to solve only the equations

\begin{equation} 
R_1 - c \tau  \: =  \: 0, 
\end{equation}

\begin{equation}
R_2 - c \bar \tau \: =  \: 0 , 
\end{equation}

\noindent to obtain $\tau ,  \: \bar \tau $ as functions of $t$, when 
suitable portions of the past trajectories ${\bf r}_1 (t) 
,  \: {\bf r}_2 (t)$ are known. Specifically, for each given $t$, 
we compute $\tau$ and $\bar\tau$ as the zeros of the functions 
$z_1 ( \tau )  \equiv || r_2 (t) - r_1 (t - \tau ) || - c 
\tau $, and $z_2 ( \bar\tau )  \equiv  || r_1 (t) - r_2 (t - \bar\tau 
) || - c \bar\tau $. The labelling has been chosen to make the 
equations below symmetrical.

We also introduce some notation to describe the velocity and 
acceleration at the retarded times: 

\begin{equation} 
{\bf v}_1 ( t )  \: = \:  {d {\bf r}_1 \over dt} ( t) , 
\quad 
{\bf v}_{1 \, \mbox r } (t)  \:  = \:  {\bf v }_1 \,  ( t - \tau ),  
\quad 
{\bf a}_{1 \,  \mbox r } (t)  \: = \: {d {\bf v}_1 \over dt} 
( t - \tau ) ,
\end{equation}

\begin{equation} 
{\bf v}_2 ( t )  \: = \:  { d {\bf r}_2  \over dt} ( t) , 
\quad
{\bf v}_{2 \, \mbox r }(t)  \: = \:  {\bf v}_2 \,  
( t - \bar \tau ),
\quad 
{\bf a}_{2 \,  \mbox r } (t)   \: = \:  
{d {\bf v}_2 \over dt} ( t - \bar \tau ) .
\end{equation}

\noindent For the vector $\bf u$ used to simplify the appearance 
of \eqrefa  {em-force} we now have the expressions

\begin{equation} 
{\bf u}_1  \: = \:  c {\bf \hat R}_1  \: - \: {\bf v}_{1 \, \mbox  r } , 
\quad \mbox {and} \quad 
{\bf u}_2  \: = \:  c {\bf \hat R}_2  \: - \: {\bf v}_{2 \, \mbox  r } .
\end{equation}

With these notations, the equations of motion take the explicit form:

\begin{equation} 
{d {\bf v}_1 \over dt}  \: = \:  \kappa  \left [ {\bf \tilde E}_2 \: 
+ \: { {\bf v}_1 \over c}  \times  ( {\bf \hat R}_2 \times 
{\bf \tilde E}_2 ) \right ] ,
\label {kappa}
\end{equation}

\begin{equation} 
{d {\bf v}_2 \over dt}  \: = \: \mu  \kappa  \left [ 
{\bf \tilde E}_1  \: + \: {{\bf v}_2 \over c}  \times  
( {\bf \hat R}_1  \times {\bf \tilde E}_1 ) \right ] ,
\label {mu}
\end{equation}

\noindent where

\begin{equation} 
{\bf \tilde E}_2   \: = \:  
{R_2  \over ( {\bf R}_2 \cdot { \bf  u}_2 )^3 } 
  \:  \: \left [ {\bf u}_2 \: (c^2 - v_{2 \, \mbox r} ^2 )   
\: \: +  \:  \:  
{\bf R}_2 \: \times  \: ({\bf u}_2 \:  \times   \:  
{\bf a}_{2 \, \mbox r} ) \right ] ,
\end{equation}

\begin{equation}
{\bf \tilde E}_1   \: = \:  
{ R_1  \over ({\bf R}_1 \cdot {\bf u}_1 )^3 } 
 \:  \: \left [ {\bf u}_1 \:  (c^2 - v_{1 \, \mbox r} ^2 )   
 \:  \: +  \:  \:  
{\bf R}_1 \: \times  \: ({\bf u}_1 \:  \times   \: 
{\bf a}_{1 \, \mbox r} ) \right ].
\label {eomend}
\end{equation}

The slight asymmetry between the two equations \eqrefa  {kappa} 
and \eqrefa  {mu} is only apparent, but is helpful as follows. 
By convention we take the mass ratio  $\mu \leq 1$, so the equation 
for the heavier particle is the one in which $\mu$ appears (and to 
calculate $\kappa$ we must use the mass of the lighter particle). For 
the purpose of computation we took  $\mu$ as the mass ratio of the 
electron to the proton, and calculated $\kappa$ accordingly. 

From the numerical point of view, the inverse of the mass ratio is 
a measure of the numerical stiffness of the problem. Thus, the problem 
is moderately stiff, and the numerical stiffness can be reduced by 
solving for a shorter time interval.

\subsection {Actual system of equations to be solved numerically}

\noindent The actual system of equations to be solved are the 12 
retarded equations,

\begin{equation}
\dot y_{[i]}    \: = \:  f_i  ( t , y_{[i]} , y_{[i]} 
(t - \tau ) ) ,
\end{equation}

\noindent where

\begin{eqnarray} 
y_{[1]}  \: = \: x_1 , \quad
& y_{[2]}  \: = \: y_1 , \quad
& y_{[3]}  \: = \: z_1 , \\ 
y_{[4]}  \: = \: x_2 , \quad
& y_{[5]}  \: = \: y_2 , \quad
& y_{[6]}  \: = \: z_2 . 
\end{eqnarray}

\begin{eqnarray} 
y_{[7]}  \: = \:  v_{1x} , \quad 
& y_{[8]} \: = \: v_{1y} , \quad
& y_{[9]} \: = \: v_{1z} , \\    
y_{[10]}  \: = \:  v_{2x} , \quad  
& y_{[11]}  \: = \:  v_{2y} , \quad
& y_{[12]}  \: = \:  v_{2z} .
\end{eqnarray}

\noindent and 
${\bf r}_1  \: = \:  ( x_1 ,  \:  y_1  ,  \: z_1 )$, 
${\bf r}_2  \: = \:  ( x_2 ,  \:  y_2  ,  \: z_2 )$, 
${\bf v}_1  \: = \:  ( v_{1x} , \:  v_{1y} ,  \: v_{1z} )$ , 
${\bf v}_2  \: = \:  ( v_{2x} , \:  v_{2y} ,  \: v_{2z} )$.

Further, 

\begin{equation} 
\dot y_{[i]}  \: = \:  f_i  \: \equiv \:  y_{[i+6]} 
\quad  \mbox {for} 
\:  1 \leq  i \leq 6 ,
\end{equation}

\noindent while for $7 \leq i \leq 12$, the functions $f_i$ are to 
be computed using the equations of motion \eqrefa  {eom}--\eqrefa  {eomend}.

\subsection {Initial functions}

\noindent The past history is prescribed as rigid two-body rotation 
around a common centre of mass, 

\begin{equation} 
{\bf r}_1 (t)  \: = \:  r_1 ( \cos  \, \omega t  ,  \: 
\sin \, \omega t , 0) , 
\end{equation}

\begin{equation} 
{\bf{r} }_ 2  (t)  \: = \:  
r_2 ( \cos  \: \omega t  ,  \: \sin \:  \omega t, 0)   ,
\end{equation}

\noindent in a coordinate frame in which the centre of mass is 
fixed. From the definition of the centre of mass, 
$m_1 r_1  \: + m_2 r_2  \: =  \:  0$, so that 
$r_e / r_p  \: = \:  1 / \mu $, with 
$r_e \: = \: r_1$ denoting the lighter particle 
according to the above convention for $\mu$. Thus, 
$r_2  \: = \: - \mu r_0$. We took $r_1  \: = \:  r_0$ = 
classical atomic radius, and $\omega$ as the corresponding angular 
velocity required to balance the Coulomb force.  We also need the 
past history of the velocities of the two particles, which is 
implicit in the above prescription. Explicitly, 

\begin{equation} 
{\bf v}_1 ( t)  \: = \:  v_1 ( - \sin  \,  \omega t ,  \:  
\cos  \, \omega t , 0 ) , 
\end{equation}

\begin{equation} 
{\bf v}_2 ( t)  \: = \:  v_2 ( - \sin  \,  \omega t ,  \:  
\cos  \, \omega t , 0 ) ,
\end{equation}

\noindent with $v_1  \: = \:  \omega r_1$, 
$v_2  \: = \:  \omega r_2 \: = \:  - \mu v_1 $. 

\noindent 

\subsection {Choice of units for computation}

\noindent At this stage a new problem arises. The numerical values 
of the quantities we have proposed to use are as follows: 
$\mu \: = \: 5.436 \times  10^{-4} $, 
$\kappa \: = \:  -2.528 \times 10^2 \: \mbox m^3 \mbox s^{-2}$, 
$r_0  \: = \:  5.3  \times 10^{-11}$~m, 
$v_0  \: = \:  \sqrt { \left ( {\kappa / r_0 } \right )}$
= $2.18 \times 10^6 \: \mbox {ms}^{-1}$, 
$\omega  \: = \:  { v_0 / r_0 }  \: = \:  
4.12 \times 10^{16} \: \mbox s^{-1}$, 
$c  \: = \:  2.99 \times 10^8 \: \mbox {ms}^{-1}$. These 
numbers, spanning 27 orders of magnitude, present too wide a range 
for reliable practical computation. 

This problem can be overcome by choosing computational units 
appropriately.\textit{The appropriate computational units are 
those in which all  quantities are represented by  numbers that 
are as close as possible to 1}. Hence, the base SI units, or units 
in which $c = 1$ are \textit{not} the most appropriate, and it is 
preferable to take length in Angstroms or  deci nano meters 
(1~dnm = $10^{-10}$~m) , and time in centi femto seconds 
(1~cfs = $10^{-17}$~s), so that the numerical values of the 
dimensioned quantities are as follows:  
$c \approx 30$~dnm/cfs, 
$\kappa \: = \: -0.02528$~(dnm)$^{3}$~(cfs)$^{-2}$, 
$r_0  \: = \:  0.53$~dnm,  
$v_0  \: = \: ~0.218$~(dnm)~(cfs)$^{-1}$, 
$\omega \: = \: 0.412$~(cfs)$^{-1}$. 
The numbers now span only 3 orders of magnitude. This choice of 
computational units however means that the numerical solutions 
can be typically computed only for a few femto seconds. This is 
nevertheless adequate for our aims. 

\subsection {The solutions}

\begin{figure}[!h]
\centerline {
\scalebox{2.0}[2.0]{\includegraphics 
		[width=12pc] {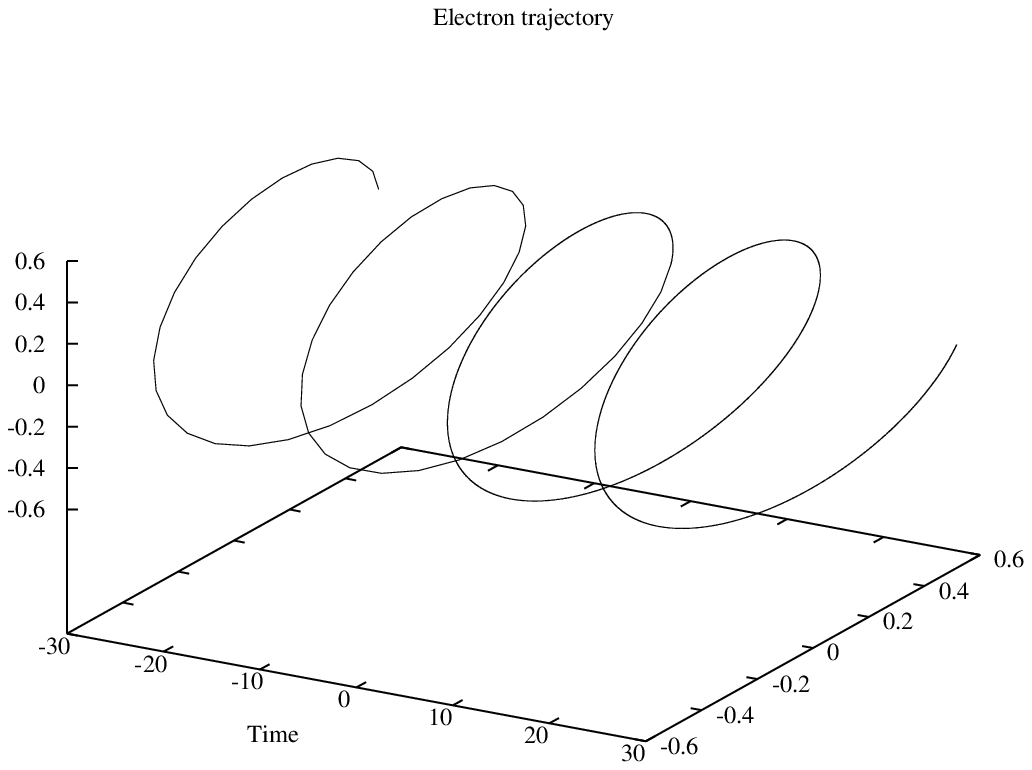}}}
\caption {\small{\it{If the past history ($t < 0 $) is prescribed as a Keplerian 
orbit, then, in the absence of radiation damping, there is no visible 
change in the electron orbit for the classical hydrogen atom.
}}}
\label {electron}
\end{figure}

\begin{figure}[!h]
\centerline {
\scalebox{2.0}[2.0]{\includegraphics 
		[width = 12pc] {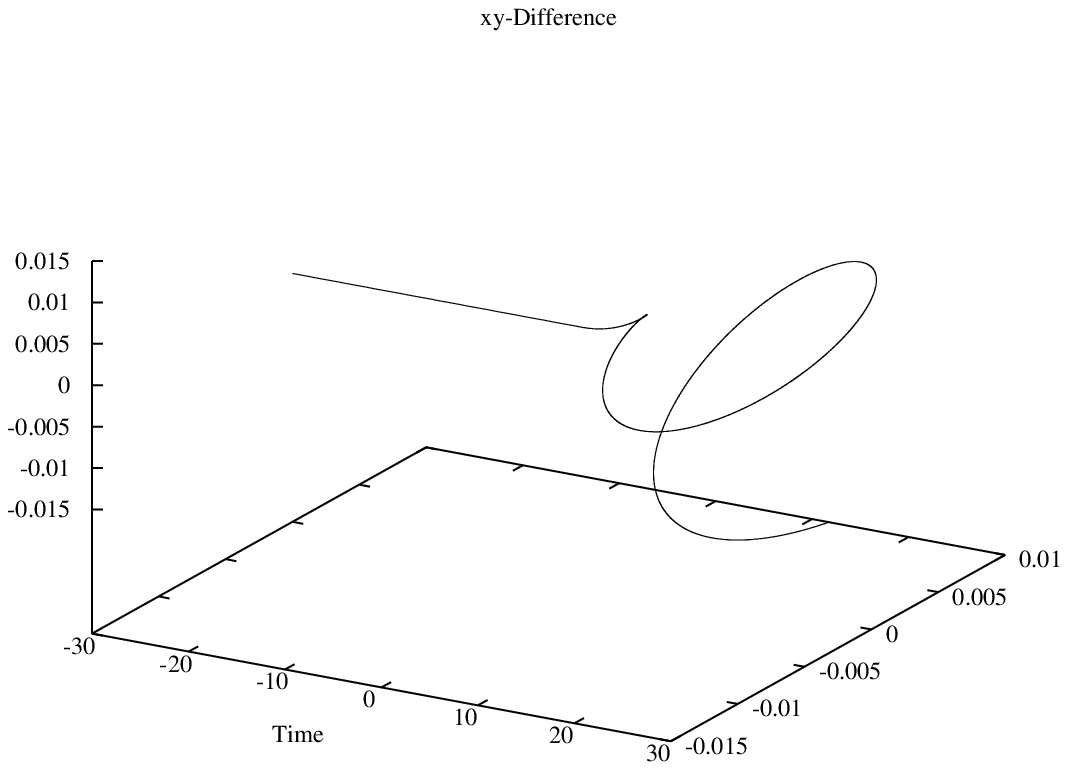}}}
\caption{\small{\it{The time evolution of the difference from the classical orbit of the 
electron. Time is in centi femto seconds, and distances are in deci 
nano meters.  The zero difference part corresponds to the prescribed 
past history. The difference that was not visible in the previous 
figure, is now seen to be large.} }
}
\label {electron2}
\end{figure}

\begin{figure}[!h]
\centerline {
\scalebox{2.0}[2.0]{\includegraphics 
		[width=12pc]{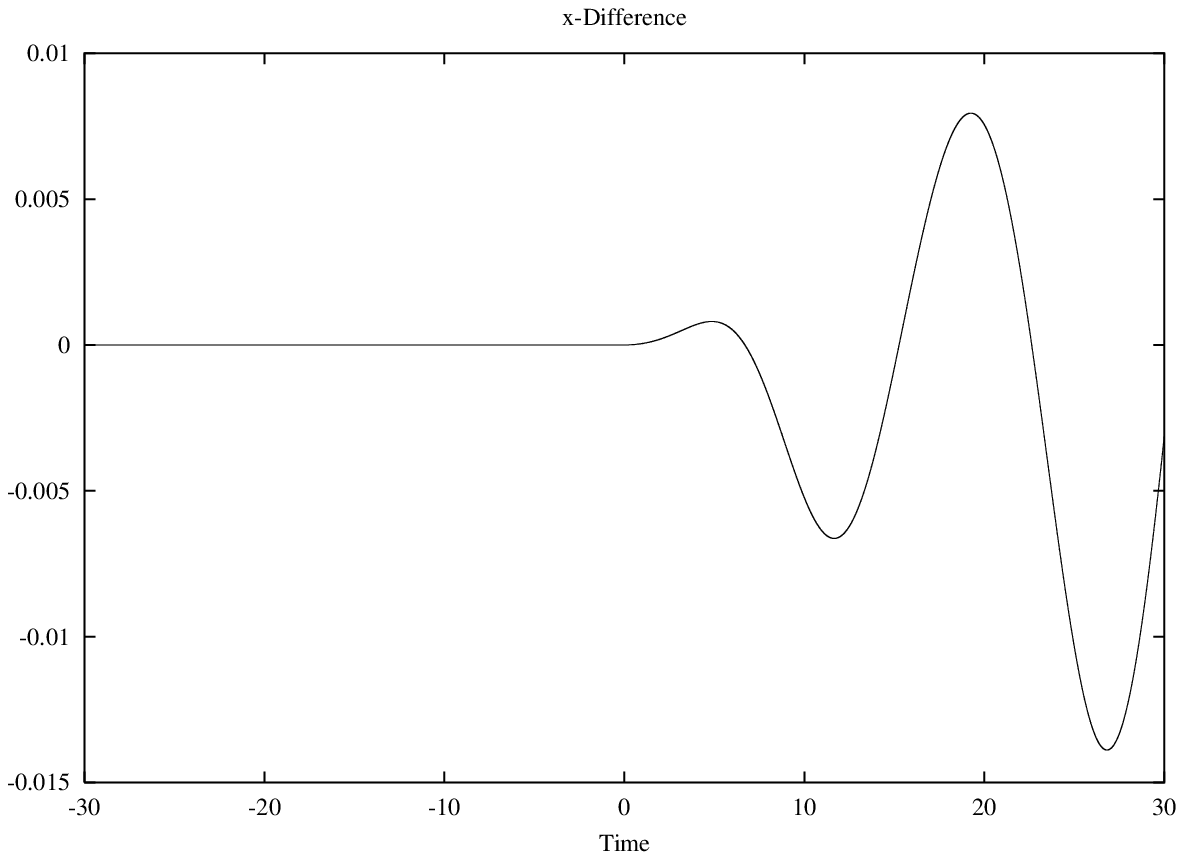}}} \par
\caption {\small{\it The time evolution of the difference of only the $x$-coordinate 
of the electron from its classical expectation. Same scale as before. 
}}
\label {electron3}
\end{figure}

\noindent The solutions were obtained using the \textsc{retard} 
program of Hairer  et al.$^{(22)}$ The results of the numerical 
computation are shown in the accompanying figures. These results 
establish that it is faulty to use a model based on the balance of 
instantaneous forces in a circular orbit, with the Coulomb force 
as in the classical Kepler problem. If past history is prescribed 
on the basis of this model, for a proton-electron pair, the 
evolution takes place as shown on the right of the $y$-axes in 
Fig.~\ref{electron}. Though the difference is not immediately 
visible, it is substantial. The differences from the classically 
expected trajectory are shown in 
Figs.~\ref{electron2}, \ref{electron3}. The differences are 
well outside the prescribed tolerances. 

\renewcommand{\textfraction}{0}

\subsection {The delay torque}

\begin{figure}[t]
\centerline {
\scalebox{2.0}[2.0]{\includegraphics
			 [width=12pc]{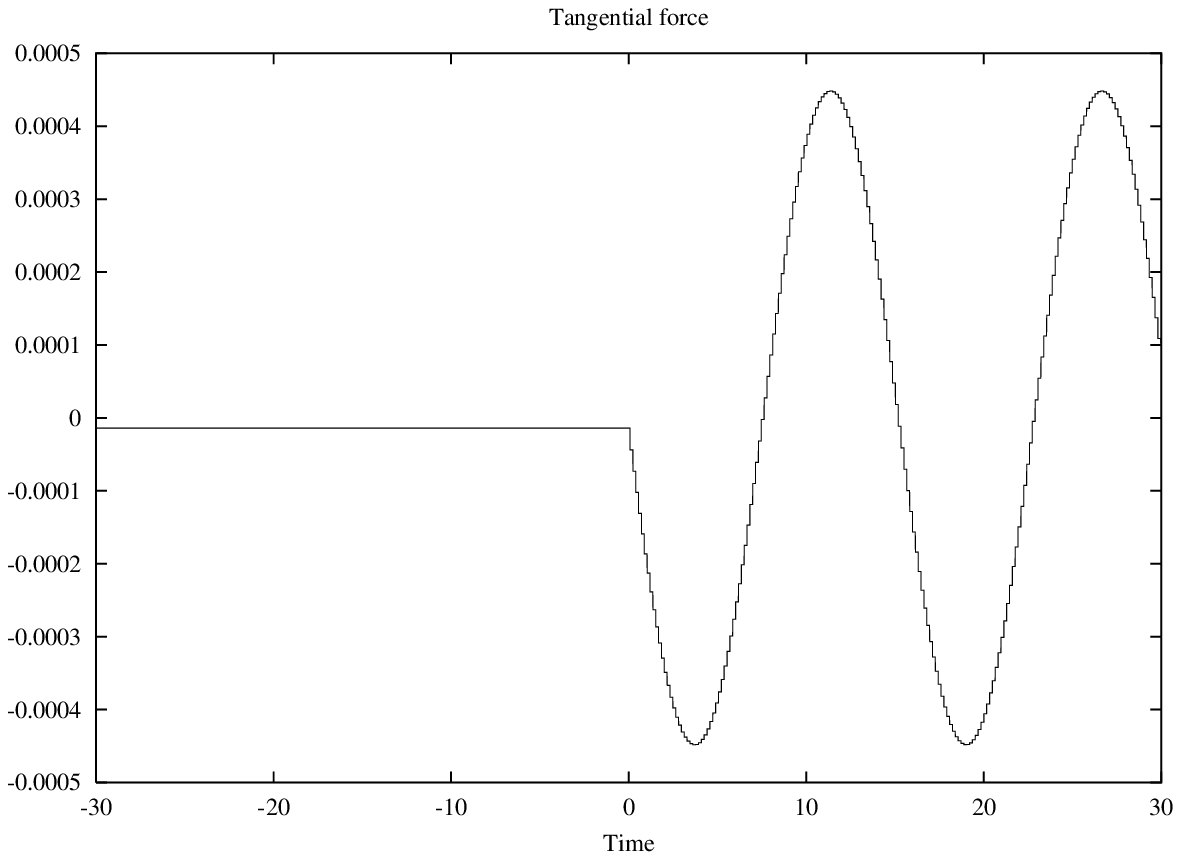}}}
\caption {\small{\it The plot shows the time variation in the tangential 
component of the force exerted by the proton on the electron. This 
component is constant during the prescribed past history of the rigid 
rotation. Subsequently, it oscillates. Thus, the electron is 
alternately accelerated and retarded. 
}}
\label{delaytorque}
\end{figure}

\noindent Apart from the purely numerical difference, there is 
a new qualitative feature. The full force acting on the electron has 
a tangential component along the direction of the electron velocity. 
Accordingly, it has a torque about the centre of mass, which torque 
initially \textit{accelerates} the electron. With the Coulomb force, 
such a tangential component of force, or a torque, is clearly 
impossible: since the Coulomb force acts instantaneously, it 
always acts along the line through the instantaneous centre of 
mass. The time evolution of this torque is exhibited in 
Fig.~\ref{delaytorque}. During the prescribed past history, 
this torque is a constant. Subsequently, it oscillates, so that 
the FDE solution oscillates about the ODE solution, and the 
entirety of the FDE orbit never strays very far from the 
entirety of the ODE orbit.    

\section {HEURISTICS}

\subsection {The need for heuristics}

\noindent Given the fundamental differences between FDE and ODE, 
it would have been astonishing if the above calculations had 
\textit{not} shown up a difference from classical expectations. 
However, the above results are based on a mathematical theory, 
which, despite its rigor, is not commonly known to physicists. 
Further, the results of a complex numerical calculation might 
well involve some hidden error or subtle instability. Accordingly, 
one gains greater confidence in the numerical results if there 
is a way to understand the results intuitively and heuristically. 
This is not easy, given that the physicist's intuition is trained 
around ODE's, that are fundamentally different from the FDE's used 
in the above calculation. However, we may proceed as follows.

\subsection {The retarded inverse square force}

\noindent For an intuitive understanding of the above result, we 
need to approximate the force \eqrefa  {em-force}  by a simpler 
force. Any instantaneous approximation, such as the Coulomb force, 
would be mathematically unsound.  Nothing we know, however, prevents 
us from approximating an FDE by a simpler FDE. That is, we may well 
approximate the full electrodynamic force by a simpler force 
\textit{while retaining the delay}. 

\begin{figure}
\centerline {
\includegraphics
		[width=12pc] {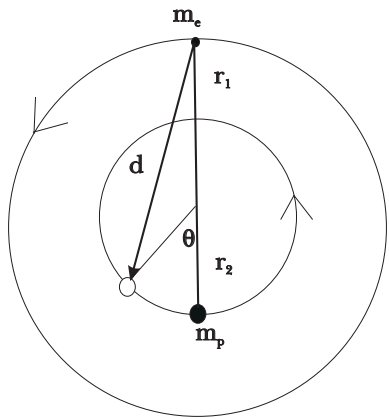}} \par
\caption {\small{\it The figure shows two charged particles $e$ and $p$, 
with masses $m_e$ and $m_p$, rotating rigidly about a 
common centre of mass $C$, in circles of radii $r_1$ and $r_2$ 
respectively, with an  assumed uniform angular velocity $ \omega$. 
The classical force exerted by $p$ on $e$ acts along the line 
joining $e$ to the `last seen' position of $p$, which 
differs from its instantaneous position by the delay angle 
$ \theta \: = \: \omega \: \delta t$, where both $ \delta t$ and 
$d$ are uniquely fixed by the condition $ \delta t \: = \: d / c$, 
$c$ being the speed of light.  
}}
\label {twobodyorbits}
\end {figure}

Accordingly, let us consider a force which varies inversely as the 
square of the \textit{retarded} distance.  The retarded distance is 
the distance from the current position of the electron to the `last 
seen' position of the proton (Fig.~\ref{twobodyorbits}).
Such a force is not physically acceptable, since it cannot 
be relativistically covariant; however, the simple expression for 
the force helps us to understand intuitively the consequences of a 
delay. 

\subsection {The delay torque}

\noindent The key point here is not the exact magnitude of the 
force, but  its \textit{direction}: for this force acts along the line 
joining the current position of the electron to the \textit{last seen} 
position of the proton. We can now immediately see the effect of even 
a tiny delay. Assume, to begin with, that the two particles are rotating 
rigidly in circular orbits about a common centre of mass. Because of 
the retardation, howsoever tiny, the last seen position of the proton 
can never be the same as its instantaneous position. Therefore, 
because of the retardation, howsoever tiny, the instantaneous force 
acting on the electron can \textit{not} pass through the instantaneous 
centre of mass. The situation for the other charged particle is similar. 
These circumstances are anathema to classical mechanics, for it is 
a well known that, under such circumstances, angular momentum cannot 
be conserved. Thus, we have the astonishing situation that, 
\textit{purely under the action of internal forces, the system 
suffers a net torque}. 

The last seen position of the proton always lags behind  its 
instantaneous position. Hence, for the situation of rigid 
rotation depicted in Fig.~\ref{twobodyorbits}, the force 
exerted by the proton on the electron has a positive component 
in the direction of the electron velocity. Hence, the force 
tends to accelerate the electron. 

We can use the retarded inverse-square force approximation also to 
heuristically try to understand the oscillation of the computed delay 
torque. The initial delay torque first accelerates both particles. 
The last-seen position of the proton starts moving towards and 
eventually falls \textit{on} the instantaneous normal line to 
the curve being traced by the electron, so that the delay torque 
(about the instantaneous centre of rotation/curvature) is zero. 
The last-seen position of the other particle (proton) then moves 
`ahead' of the instantaneous normal line to the curve traced by 
the electron, where  `ahead' and `behind' is with respect to 
the direction in which the curve is being traced. The delay 
torque changes sign and starts decelerating the electron. If 
we neglect radiation damping, angular momentum is conserved 
on the average. This is similar to what the numerical computation 
with the full electrodynamic force showed. 

\subsection {Radiation damping and stability}

\noindent There is an old argument to the effect that the 
introduction of radiation damping makes the classical hydrogen 
atom unstable. This argument is based on the Coulomb-force 
formulation of the electrodynamic 2-body problem. For the full 
electrodynamic force, the effect of introducing radiation 
reaction is no longer obvious, because of the existence of a 
delay torque which initially \textit{accelerates} the electron. 

There are three difficulties here. First the exact form of the 
radiation damping term is not quite clear. Dirac's$^{(23)}$ 
argument that the 5th and higher-order terms are too complicated 
for `a simple thing like the electron' is an interesting argument, 
but not entirely convincing. Secondly, this author has pointed 
out that Dirac's derivation of radiation reaction approximates 
a FDE by a ODE using a `Taylor' expansion in powers of the delay, 
and the mathematically suspect nature of this approximation is 
reinforced by the above calculation. Finally, putting in 3rd 
order radiation reaction may be expected to result in a numerically 
ill-conditioned problem, and appropriate codes for this are 
yet to be developed. (The \textsc{dopri 4-5 }code used 
implicitly in \textsc{retard} is not appropriate for stiff 
ODE's.)

Since the issues involved are complex, it is helpful to have 
an initial estimate about the outcome of such a rigorous 
calculation.  For the purposes of such an initial estimate, 
about the effect of radiation damping, we again use the simple 
heuristic case of the retarded inverse square force, to 
determine whether there can be a balance of forces between 
the delay torque and 3rd order radiation damping.

To this end, consider two particles interacting through the 
retarded inverse square force. The force on the electron 
exerted by the proton is given by 

\begin{equation} 
{\bf F} \: = \:  { e^2   \over   d^3 }  \:  {\bf d}  .
\end{equation}

\noindent The force acts in the direction of the 3-vector 
$\bf d$ ,  along which the proton is `last seen' by the electron. 
The 3-vector $\bf d$  may be represented by 

\begin{equation} 
{\bf d} \:  =  \:  {\bf r}_p (t - \delta t )  \: - \:  
{\bf r}_e (t) ,
\end{equation}

\noindent where ${\bf r}_p ( t)$, and ${\bf r}_e ( t )$ 
denote respectively the instantaneous position vectors of the 
proton and electron, respectively, at time $t$, and $\delta t$ 
is the delay, so that  $ {\bf r}_p (t - \delta t )$ is the 
`last seen' position of the proton.

Assuming that the two particles are in rigid rotation with 
constant angular velocity $\omega$, and referring back to 
Fig.~\ref{twobodyorbits}, we have, in 3-vector notation,  

\begin{eqnarray}
{\bf r}_e  \:  & = & \:  r_1 [ \cos \, \omega t \, { \bf \hat \imath}
  \: +  \:  \sin \, \omega t \, {\bf \hat \jmath}  ] , \\
- {\bf r}_p  ( t -  \delta t )  \:  & =  & \: r_2 [ 
\cos  \, \omega (t - \delta t)   \, { \bf \hat \imath}  \: + 
 \:  \sin \,  \omega (t - \delta t )   \, {\bf \hat \jmath}  ] \\
 & \approx & r_2 \,  [ \cos \,  \omega t \: + \:  \omega \delta t  
 \: \sin  \,  \omega t ]  \,  { \bf \hat \imath} \: \nonumber \\
& & {} + \:  r_2 \, 
 [ \sin \,  \omega t  \: - \:  \omega \delta t  
 \: \cos  \, \omega t ] \,  { \bf \hat \jmath }   \, ,
\end{eqnarray} 

\noindent assuming that $\omega \delta t$  is small. This last 
assumption is justified since, for classical rigid rotation, 
$\omega~=~{v_0 / r } $ , while 
$\delta t~\sim~r / c $,  so that 
$\omega \delta t~\sim~v_0 / c~\sim~0.01$.  

Hence, the delay torque on the electron is given by

\begin{equation} 
{\bf r}_e  \,  \times  \:  {\bf F}  \: = \:  { e^2 \over d^3 }  
 \:  {\bf r}_e \times  {\bf d}  \: = \:   
{ e^2 \over d^3} r_1 r_2  \, \omega \delta t  \,  {\bf \hat k}   .
\end{equation}

\noindent To a first approximation, we may use $d~\sim~r_1  \: +
\: r_2  \: = \:  (1  \: +  \: \epsilon ) r_1$, where 
$\epsilon \: = \:  { m_e /  m_p} $, is the ratio of the 
electron to proton mass. Accordingly, 
$r_1 r_2  \: = \:  \epsilon r_1^2 $, and, further using 
$\delta t~\sim~{d / c}$, the 
expression for the delay torque simplifies to

\begin{equation} 
T_{\mbox {delay}} \: = \: { e^2 \over c } 
\cdot { \epsilon \over (1 + \epsilon )^2 }  \, \cdot  \,    
\omega  \:  {\bf \hat k} \:  \: \approx  \:  
{\epsilon e^2  \over c} \omega {\bf \hat k} .
\end{equation}

On the other hand, for the `average' force due to (3rd order) 
radiation reaction, we have the well-known formula$^{(24)}$

\begin{equation} 
{\bf F}_{\mbox {rad}}   \: = \:  {\mu_0 q^2 \over 6 \pi c }  
{\bf \dot a}_e  \: = \: {1 \over 4 \pi \epsilon_0 } 
\cdot {2 \over 3} {q^2  \over c^3} {\bf \dot a}_e  \: = \:  
{2 \over 3} {e^2 \over c^3} {\bf \dot a}_e  .
\end{equation}

\noindent For the case under consideration,  

\begin{equation} 
{\bf \dot a}_e  \: = \:  { d {\bf a}_e \over dt }
 \: = \:  - \omega^2 {\bf v}_e  . 
\end{equation}

\noindent Hence, the torque due to radiation damping

\begin{equation} 
T_{\mbox {rad }}  \: = \:  {\bf r}_e  \:  \times  \: 
{\bf F}_{\mbox {rad}}  \: = \:  {2 \over 3} {e^2 \over c^3}  
\cdot  \:  - \omega ^2 \cdot {\bf r}_e \times  
 \: {\bf v}_e  \: = \:  -  \,  \: {2 \over 3} {e^2 \over c^3}  \, 
\omega^3  \, r_e^2  \, { \bf \hat k}  . 
\end{equation}

The two torques are oppositely directed. Thus, a necessary condition 
for conservation of angular momentum is

\begin{equation} 
T_{\mbox {delay}}  \: + \:  T_{\mbox {rad}}  \: = \:  0,
\end{equation} 

\noindent which gives

\begin{equation} 
{e^2 \over c} \cdot {\epsilon \over ( 1 + \epsilon )^
2 } \cdot \omega  \: = \: { 2 \over 3} {e^2 \over c^3}  \:  
\cdot \omega^3 r_e^2  ,
\end{equation}

\noindent or

\begin{equation} 
\omega^2 \: = \:  {3 \over 2} {\epsilon \over (1+ \epsilon )^
2}  
\cdot {c^2  \over r_e^2} \:  \approx   \:  
{3 \over 2} {\epsilon c^2  \over r_e^2 } .
\end{equation}

\noindent Once the constraint of rigid rotation is removed, many 
such equilibrium solutions may well exist, with or without radiative 
damping. 

We note, incidentally, this condition for conservation of angular 
momentum is quite different from the classical condition for balance 
of forces, which requires $\omega^2  \: = \:  e^2 / r_e^3 $. Both 
conditions can hold simultaneously only for one value 
of $r$ which is smaller than the Bohr radius.

Thus, further investigations are required to determine the exact 
effects of radiative damping, and it was prematurely concluded that 
radiative damping makes the classical hydrogen atom unstable.  

\subsection {Discrete spectrum and FDE's}

\noindent The \textit{discreteness} of the observed  hydrogen spectrum 
has been regarded as another compelling argument against classical 
electrodynamics. Solutions of FDE's can, however, admit an infinite 
discrete spectrum.

Thus, for example, consider the \textit{retarded harmonic oscillator}, 
given by the linear, second order, retarded FDE 

\begin{equation} 
\ddot x (t)  \: = \:  -x (t-1) , 
\label {rhm}
\end{equation}

\noindent It is easy to see that a function of the form

\begin{equation} 
x (t)  \: = \:  e^{z_k  \,  t} , 
\label {exponent}
\end{equation}

\noindent for complex $z_k $, is a solution of the equation 
\eqrefa  {rhm} if and only if $z_k $ is a  solution of the\textit{
quasi}-polynomial equation 

\begin{equation} 
z^2 e^z  \: = \: - 1 .
\label {quasi}
\end{equation}

It is equally easy to see that the quasi-polynomial equation 
\eqrefa{quasi} has \textit{infinitely many} complex solutions 
$z_k  \: = \:  x_k  \: \pm \:  i  \,  y_k $, and it is 
known$^{(25)}$ that the roots are discrete, with no cluster 
point, and that the large magnitude roots are given asymptotically 
by the approximate expression  

\begin{equation} 
y _ k  \: = \:  2 k  \pi  \: +  \:  \epsilon_1 (k),
\end{equation}

\begin{equation} 
x_k  \: = \:  - \ln  \, y_k  \: + \:  \epsilon_2 (k) ,
\label{roots}
\end{equation}

\noindent where $\epsilon_1 (k)  \:  \rightarrow 0$, and 
$\epsilon_2 (k)  \:  \rightarrow 0$ as $| z_k |  \: \rightarrow \infty $. 

(Had we dropped the retardation from \eqrefa{rhm}, and applied 
the same procedure instead to the usual harmonic oscillator, given 
by the ODE  $\ddot x (t)  \: = \:  -x(t)$, that would have led, in the 
well-known way, to a quadratic polynomial $z^2  \: = \:  -1$ with 
exactly one pair of complex conjugate roots: $i$, and $-i$.)

Since the equation \eqrefa {rhm} is linear, any linear combination 
of these infinitely many oscillatory solutions of the form 
\eqrefa {exponent} is again a solution. Setting aside questions of 
convergence, it is clear that any \textit{finite} linear combination 
of the form

\begin{equation} 
x ( t)  \: = \:  \sum_{k = 1} ^ n  \:  a_k  \, e^{z_k  \, t },
\label{general soln}
\end{equation}

\noindent with $z_k $ given by \eqrefa {roots}, say, 
is also a solution of \eqrefa {rhm}. That is, in physical terms, 
the retarded harmonic oscillator, governed by the simple linear FDE 
\eqrefa {rhm}, exhibits an infinite spectrum of discrete frequencies. 
The general solution is a convergent linear combination of 
oscillations at an \textit{infinity} of discrete (`quantized') 
frequencies. As in quantum mechanics, to determine a unique 
solution one needs to know an `initial' function. If the initial 
function is prescribed arbitrarily, we can expect discontinuities. 

The above considerations remain valid for any linear FDE with constant 
coefficients and constant delay.$^{(26)}$ The locally linear 
approximation$^{(27)}$ suggests that such `quantization' is also 
to be expected for the non-linear FDE's of the electrodynamic 
2-body problem. [For the FDE \eqrefa {FDE} this approximation may be 
obtained simply by replacing $f$ by its first-order `Taylor' 
expansion, and then freezing the values of the delays $t_r$, in a 
neighborhood of the point at which we want to approximate the 
solution.]

Mere discreteness of the spectrum does not, of course, mean that the 
FDE formulation, using retarded electrodynamics, will correctly 
reproduce the actual spectrum of the hydrogen atom. Equally, mere 
discreteness of the spectrum was not an adequate argument 
\textit{against} the classical hydrogen atom, which was prematurely 
rejected  on the basis of the ODE formulation of the 2-body 
problem of electrodynamics. Though a further investigation of 
FDE's may still lead to a rejection of the classical hydrogen 
atom for the right reasons, it  could, on the other hand, well 
help to clarify the `missing link' between classical and quantum 
mechanics, as in the structured-time interpretation of 
quantum mechanics.$^{(28)}$

\section {CONCLUSIONS}

\noindent 1. The Coulomb force is not a good approximation to the 
full electrodynamic force between moving charges. 

\noindent 2. The classical hydrogen atom was prematurely rejected 
on the basis of the ODE formulation of the electrodynamic 2-body 
problem.

\noindent 3. The \textit{n}-body problem involving electrodynamic 
interactions---as formulated, for example, in current software for 
molecular dynamics---needs to be reformulated using FDE's.

\subsection {\textbf{Acknowledgments}}

\noindent The author is grateful to Profs G. C. Hegerfeldt 
(G\"{o}ttingen), Huw Price (Sydney), David Atkinson (Groningen), 
H. D. Zeh (Heidelberg), Dennis Dieks (Utrecht), Jos Uffink 
(Utrecht), C. J. S. Clarke (Southampton), P. T. Landsberg 
(Southampton), Amitabh Mukherjee (Delhi), and Mr Suvrat Raju (Harvard) 
for discussions, and to an anonymous referee for comments that have 
greatly helped to improve the presentation of the paper.

\pagebreak    

{\bf NOTES AND REFERENCES}

1. C. K. Raju, \textit{Time: Towards a Consistent Theory}, 
Kluwer Academic, Dordrecht, 1994. (Fundamental Theories of Physics, 
vol. 65.)

2. Preliminary versions of aspects of this paper have 
been circulating for some time, having been presented and discussed 
at various conferences and  lectures over the past several years, 
e.g. ``Simulating a tilt in the arrow of time: preliminary results,'' 
Seminar on \textit{Some Aspects of Theoretical Physics}, Indian 
Statistical Institute, Calcutta, 14--15 May 1996,  ``The electrodynamic 
2--body problem and the origin of quantum mechanics.'' Paper 
presented at the International Symposium on \textit{Uncertain Reality}, 
New Delhi, 5--9 Jan 1998, ``Relativity: history and history 
dependence.'' Paper presented at the \textit{On Time} Seminar, British 
Society for History of Science, and Royal Society for History of Science, 
Liverpool, August 1999. ``Time travel,'' invited talk at the 
International Seminar, \textit{Retrocausality Day}, University of 
Groningen, September 1999, and in talks at the Universities of 
Southampton, Utrecht,  Pittsburgh, etc.

3. Some attempts have been made to study the 2-body problem 
in 1 and 2-dimensions, and some approximation procedures have been 
suggested for three dimensions, but none of these are critically 
relevant to the question at hand. For a quick review see 
C. K. Raju, ``Electromagnetic Time,'' chp. 5b in 
\textit {Time: Towards a Consistent Theory}, cited earlier. For the  
1-body problem see C. F. Eliezer, 
\textit{Rev. Mod. Phys.} \textbf {19} (1947) p. 147 ; 
G. N. Plass, \textit{Rev. Mod. Phys.} \textbf{33}, (1961) pp. 37--62.  
For the 2-body problem, 
see J. L. Synge, \textit{Proc. R. Soc.} \textbf{A177} (1940) 
pp. 118--139, 
R. D. Driver, \textit{Phys. Rev.} \textbf{178} (1969) pp. 2051--57, 
D. K. Hsing, \textit{Phys. Rev.} \textbf{D16} (1977) 
pp. 974--82, A. 
Schild, \textit{Phys. Rev.} \textbf{131} (1963) p. 2762, 
C. M Anderssen and 
H. C. von Baeyer, \textit{Phys. Rev.} \textbf{D5} (1972) p. , 802, 
\textit{Phys. Rev.} \textbf{D5} (1972) p. 2470, R. D. Driver, 
\textit{Phys. Rev.} \textbf{D19} 
(1979) p.  1098, L. S. Schulman, \textit{J. Math. Phys.} 
\textbf{15} (1974) 
pp. 205--8, K. L. Cooke and D. W. Krumme, 
\textit{J. Math. Anal.  Appl.} 
\textbf{24} (1968) pp. 372--87, H. Van Dam and E. P. Wigner, 
\textit{Phys. Rev.} \textbf{138B} (1965) p. 1576, \textbf{142} 
(1966) p. 838. 

4. J. Andrew McCammon and Stephen C. Harvey, \textit{Dynamics 
of Proteins and Nucleic Acids}, Cambridge University Press, 1987, 
p. 61.

5. David J. Griffiths, \textit{Introduction to Electrodynamics}, 
Prentice Hall, India, 3rd ed., 1999,  p. 435, eq. 10.46. Cf.  J. D. 
Jackson, \textit{Classical Electrodynamics}, 3rd ed., John Wiley, 2001. 
Griffiths' book is more convenient for our purpose.

6. cf. Griffiths, \textit{Electrodynamics}, 3rd ed., equns. 
10.65, 10.66 and 10.67, pp. 438--39.

7. usually called just the Lorentz force law. 

8. Griffiths, \textit{Electrodynamics}, 3rd ed., 
p. 439, eq. 10.67. 

9. Griffiths, \textit{Electrodynamics}, 3rd ed., 
p. 421. O. L. Brill and B. 
Goodman, \textit{Amer. J. Phys.} \textbf{35}, 1967, p. 832.

10. C. K. Raju, ``Electromagnetic Time,'' chp. 
5b in \textit{Time: Towards a Consistent Theory}, cited earlier.

11. L. E. El'sgol'tz, \textit{Introduction to the Theory 
of Differential Equations with Deviating Arguments}, trans. R. J. 
McLaughlin, Holden-Day, San Francisco, 1966, pp. 13--19. R. D. 
Driver, \textit{Introduction to Differential and Delay Equations}, Springer, 
Berlin, 1977.

12. C. K. Raju, ``Electromagnetic Time,'' chp. 
5b in \textit{Time: Towards a Consistent Theory}, cited earlier.

13. C. K. Raju, \textit{Time: Towards a Consistent Theory}, 
cited earlier.

14. An earlier method suggested by Synge also assumes 
instantaneity, but it has the advantage that it can actually be implemented. 
This implementation will be considered in subsequent articles. See, 
J. L. Synge, ``The electrodynamic 2-body problem,'' \textit{J. R. 
Soc.} \textbf{A177} (1940) pp. 118--139. 

15. El'sgol'tz, \textit{Differential Equations with Deviating 
Arguments,} pp. 13--19. Driver, \textit{Differential and Delay Equations}.

16. E. Hairer, S. P. Norsett, and G. Wanner, \textit{Solving 
Ordinary Differential Equations}, Springer Series in Computational 
Mathematics, Vol. 8, Springer, Berlin, 1987. Revised ed. 1991. 

17. El'sgol'ts, \textit{Differential Equations with Deviating 
Arguments}, p. 9. 

18. Driver, \textit{Differential and Delay Equations}.

19. C. A. H. Paul, ``A user guide to \textsc{archi}: an explicit 
Runge-Kutta code for solving delay and neutral differential equations,'' 
Numerical Analysis Report No. 283, Department of Mathematics, University 
of Manchester, 1995. 

20. D. R. Will\'{e} and C. T. H. Baker, ``The tracking 
of derivative discontinuities in systems of delay differential equations,''  \textit{Appl. 
Num. Math.} \textbf{9} (1992) pp. 209--222.

21. Griffiths, \textit{Electrodynamics}, 3rd ed., equns. 7.40, 
and 11.80, pp. 326 and 467.

22. E. Hairer, S. P. Norsett, and G. Wanner, \textit{Solving 
Ordinary Differential Equations}, cited above. (The above 
figures used the newer  version of \textsc{retard}.)

23. P. A. M. Dirac, ``Classical theory of the radiating 
electron,'' \textit{Proc. R. Soc. A} \textbf{167} (1938) p. 148. 
Also F. Rohrlich, \textit{Classical Charged Particles}, 
Addison-Wesley, Reading,  Mass, 1985, p. 142.

24. Griffiths, \textit{Electrodynamics}, 3rd ed., \S~11.80, 
p. 467.

25. El'sgol'tz, \textit{Differential 
Equations with Deviating Arguments},  pp. 32--33. 

26. El'sgol'tz, \textit{Differential Equations with Deviating 
Arguments}, pp. 28--41. 

27. C. K. Raju, ``Simulating a tilt in the arrow of 
time,'' cited earlier.  Such an approximation is \textit{a priori} 
plausible, though there is, at present, no formal proof of its validity.

28. C. K. Raju, ``Quantum-Mechanical Time,'' chp. 
6 in \textit{Time: Towards a Consistent Theory}, cited earlier.

\pagebreak

\end {document}